\documentclass[review]{elsarticle}

\usepackage{hyperref}
\usepackage{color}
\usepackage{rotfloat}

\journal{Journal of \LaTeX\ Templates}

\begin{document}

\topmargin=+1.0in

\begin{frontmatter}

\title{Prediction of Influenza B Vaccine Effectiveness from Sequence Data}

\author[sspb]{Yidan Pan}
\author[bioe,sspb,phys]{Michael W.\ Deem\corref{mycorrespondingauthor}}
\cortext[mycorrespondingauthor]{Corresponding author}
\ead{mwdeem@rice.edu}

\address[sspb]{Systems, Synthetic, and Physical Biology,
Rice University 6100 Main St, Houston, TX  77005}
\address[bioe]{Department of Bioengineering,
Rice University 6100 Main St, Houston, TX  77005}
\address[phys]{Department of Physics \& Astronomy,
Rice University 6100 Main St, Houston, TX  77005}

\begin{abstract}
Influenza is a contagious respiratory illness that
causes significant human morbidity and mortality,
affecting 5-15\% of the population in a typical epidemic 
season.  Human influenza epidemics are caused by types A and B, with
roughly 25\% of human cases due to influenza B.
Influenza B is a single-stranded RNA virus with a high mutation
rate, and both prior immune history
and vaccination put significant pressure on the
virus to evolve.
Due to the high rate of viral evolution, the 
influenza B vaccine  component of the annual influenza vaccine is updated,
roughly every other year in recent years.
To predict when an update to the vaccine is needed, an estimate of expected
vaccine effectiveness against a range of viral strains is required.
We here introduce a method to measure antigenic distance between the influenza B
vaccine and circulating viral strains. The measure correlates well with
effectiveness of the influenza B component of the annual vaccine in 
humans between
1979 and 2014. We discuss how this measure of antigenic distance may be
used in the context of annual influenza vaccine design and prediction of
vaccine effectiveness.
\end{abstract}


\end{frontmatter}

\section{Introduction}

Influenza is a highly contagious respiratory illness that infects all ages.
 It occurs seasonally 
worldwide, causing substantial morbidity and mortality \cite{bridges2003prevention}.
In humans, influenza A/H3N1, A/H1N1, and B are the main causative agents.
 In each
influenza season, influenza A generally spreads widely. Influenza B
tends to be regionally dominant, causing seasonal epidemics every 2--4 years
\cite{thompson2003mortality,belshe2010need}. 
The major molecular factors of influenza accessible to the
immune system are hemagglutinin and neuraminidase. 
Entry into host
cells requires hemagglutinin, which is also the main target of neutralizing
antibodies \cite{ada1986immune,skehel2000receptor}. 
To protect against influenza infections, the World
Health Organization (WHO) recommends the trivalent seasonal influenza
vaccine, which includes the two main influenza type A strains, H1N1
and H3N2, and an influenza B strain from
the Yamagata or Victoria
lineages \cite{WHOeachyearIB}.
Hemagglutinin undergoes
antigenic drift because of selective pressure, and this
drift reduces vaccine effectiveness. 

Influenza B virus is less prevalent than influenza A, 
and the morbidity and mortality associated
with influenza B are often perceived to be
lower than those caused by influenza A. 
Nonetheless, 
influenza B has attracted researchers'
attention in recent years. 
An early modeling study with a
two-tiered phylodynamic model 
for influenza B compared the evolution of A/H3N2, A/H3N8, and B \cite{Koelle}.
In some recent studies, influenza
A and B infections resulted in similar morbidity and mortality in
hospitalized adults \cite{su2014comparing}, and also caused similar
clinical characteristics in outpatients \cite{irving2012comparison}.
These results indicate that influenza B virus can cause infections as severe
as those caused by
influenza A. 

Influenza B is comprised of two distinct evolutionary 
lineages: B/Yamagata/16/88-like and B/Victoria/2/87-like. These two
lineages have co-circulated since the 1980s \cite{rota1990cocirculation,mccullers2004multiple}.
The lineage that predominates can change yearly. In the USA, there were 5 changes of the predominant lineage in the 10
seasons between 2001 and 2010 \cite{dolin2013quadrivalent}, often resulting in a
mismatch between the vaccine strain and the
dominant circulating strain. 
There is limited to no protection
of a vaccine for one lineage against a viral
strain from the other lineage.
Moreover, the protection of a vaccine for one lineage against
viruses within that same lineage is not uniform, as it
depends on the antigenic distance between the vaccine
and virus.
Compared to influenza A, there are
relatively few studies of
the relation between
antigenic distance and vaccine effectiveness for
influenza B.

We here introduce a method to estimate
antigenic distance between the influenza B
vaccine and circulating viral strains. Epitope regions
were mapped from a standard influenza A/H3N2 virus.
Additional amino acid sites that exhibited high entropy
in influenza B sequences sampled from the human population
were added to the epitope regions.
Note that while the binding of
a specific antibody in a specific individual 
against a specific antigen
is typically dominated by 5--6 contacting amino acid residues in the
antigen, here we consider
 typical regions of antibody
binding among a distribution of $10^8$ antibodies and a distribution of  
$7 \times 10^9$ people and against a mixture of antigens within a
given influenza B strain.  This is the reason for the rather large
size of the generalized epitopes A-E in the hemagglutinin  protein
considered here.
We here define an estimate of
antigenic distance between the vaccine strain and
the dominant circulating strain as $p_{{\rm epitope}}$,
the fraction of amino acid site substitutions between the vaccine and circulating
virus lineage in the dominant epitope region 
of HA1 \cite{gupta2006quantifying}.
We will show that this measure of
 antigenic distance 
correlates well with effectiveness of the influenza
B component of the annual vaccine between 1979 and 2014.

To additionally detect the emergence of
new influenza strains in the human population, 
we built a dimensionally reduced space using the multidimensional
scaling method. This reduced space allowed for the
visualization of the evolution of influenza B. 
In this representation, the emergence of new vaccine strain
clusters is apparent. In particular, clades appear in
this representation at antigenic distances
sufficiently far from current dominating strains that
immune recognition is ineffective.  

This study provides a method to estimate 
influenza B antigenic distance and vaccine effectiveness.
The measure of antigenic distance introduced here, which
provides a novel tool for prediction
of vaccine effectiveness in influenza B,
may be used in the context of annual influenza vaccine design.

\section{Methods}

\subsection{Sequence Data}

Influenza B hemagglutinin protein sequences from human hosts in all
regions were downloaded from the Influenza Virus Database of the National
Center for Biotechnology Information (NCBI) \cite{bao2008influenza}.
Data were collected for the years 1979 to 2014.
Only data containing
the entire HA1 sequence were used.

\subsection{Vaccine Strain and Circulating Strain}

For each season, and for each epidemiological study, we determined
the relevant vaccine strain and the relevant dominant circulating strain.
For
those studies that specified the vaccine strain used and the dominant
circulating strain observed, we used those sequences. 
For the studies conducted in the USA that did not specify the strains, 
we used the vaccine strain and dominant circulating strain identified in the Center for Diseases Control and Prevention (CDC) 
''Morbidity and Mortality Weekly Report''.
For the remaining
studies, we used the vaccine strain and dominant circulating strain
identified in the annual WHO reports on ''recommendations on the composition
of influenza virus vaccines.''
We note that the vaccine and dominant
circulating strains changed from year to year and from Northern to
Southern Hemisphere studies.

\subsection{Sequence Alignment}

In each season, and in each  epidemiological study, we determined whether
the vaccine strain was a Yamagata or Victoria lineage strain.
To do this, we first determined the lineage to which the vaccine strain belonged
from the WHO report \cite{WHOeachyearIB}.
 If the report was inconclusive, we used the lineage
assigned by the Influenza Virus Database. If the database was inconclusive
as well, typically only for the sequences from early years, we used
the phylogenetic tree calculated by the maximum likelihood method
of MEGA 6.06 \cite{tamura2013mega6} to determine the lineage to which
a strain belonged. 
For each lineage, we used a standard crystal structure of HA1 to align
the strains. For the Yamagata-like lineage, we used B/Yamanashi/166/1998 
(PBD ID: 4M44). For the Victoria-like lineage, we used B/Brisbane/60/2008 
(PBD ID: 4FQM). All protein sequences from a given Northern or Southern
Hemisphere influenza season were aligned to the sequence determined
by this procedure. The sequences were numbered with reference to these
standards, with the first amino acid site of each model amino acid site 1. The Victoria
sequence is one amino acid longer than the Yamagata model, and multi-sequence
alignment using PRANK \cite{loytynoja2010webprank} showed a gap at
amino acid site 163 in the Yamagata model.

The crystal structures of the two models are similar, with average
root mean square deviation between them of 0.3671 \AA .There are
six missing amino acid sites in the crystal structure of 4FQM and three missing
amino acid sites in the structure of 4M44. These amino acid sites are in the tail
region. They are far away from the epitope region and are not included
in root mean square deviation calculations.

\subsection{Epitope Mapping}

We here mapped the epitopes of both Yamagata-like and Victoria-like
influenza B virus from the 5 epitopes of the H3 subtype of influenza
A virus  \cite{wilson1990structural,wilson1981structure,Epitope2,ISD,gupta2006quantifying,deem2009epitope,pan2011novel}
using PRANK, which allows for gaps in the
alignment and takes evolutionary
distances into account \cite{loytynoja2010webprank}.
We note that various modifications of these epitopes have been proposed,
which range from relatively minor additions of amino
acid sites \cite{entropy}, to dramatic
reassignments \cite{ref47}.  The latter is discussed in
our previous study \cite{xi2015H3}, and we note here that measures of antigenic
distance based upon it
correlate less well with A/H3N2 vaccine effectiveness in humans
than do those based upon the original epitope definition.

 We used the
sequence and numbering of A/California/7/2004 for the reference, as
in the H1N1 mapping studies \cite{deem2009epitope,pan2011novel}.
If an epitope amino acid site in A/California/7/2004 aligned to a gap in the
influenza B model, that epitope amino acid site was deleted in the influenza
B model. If an epitope amino acid site in A/California/7/2004 aligned to multiple
amino acid sites in the influenza B model, because of a gap in A/California/7/2004
adjacent to the epitope amino acid site, then further analysis was undertaken.
The amino acid site before the gap and all the amino acid sites in the gap in the
influenza B model were considered as potential epitope amino acid sites. The
amino acid sites with the highest seasonal entropy, as described more fully
in the Supplemental Information,
 were selected as the final epitope amino acid sites.
Numbers of sequences per year are shown in Table S1.

\subsection{Antigenic Distance between the Vaccine Strain and the Dominant 
Circulating Strain}

\label{method:antigenic_distance} The antigenic distance between
the vaccine strain and the dominant circulating strain was measured
by $p_{{\rm epitope}}$, the fraction of amino acids
in the dominant epitope region of HA1 that
differ between the vaccine and circulating
strain \cite{gupta2006quantifying}. We additionally considered several
alternative definitions of antigenic distance. The first is $p_{{\rm all~epitopes}}$,
the fraction of amino acid substitutions in the entire set of all
five epitope regions. The second is $p_{{\rm sequence}}$, the fraction
of amino acid substitutions in the entire HA1 sequence. The third
is $p_{{\rm two~epitopes}}$, introduced in this study, which is the
fraction of amino acid substitutions in the two epitopes with the
highest individual amino acid substitution fractions. The model, Yamagata
or Victoria, used in these calculations was that of the lineage to
which the vaccine strain belongs for the influenza season under consideration.
If there was a lineage mismatch when the vaccine strain was Victoria,
the gap to which amino acid site 163 mapped 
was considered as a amino acid site substitution.
If there was a lineage mismatch when the vaccine strain was Yamagata, amino acid
site 163 of the dominant strain, which mapped to a gap on the vaccine,
was excluded from the substitution computation.

\subsection{Estimation of Vaccine Effectiveness}

Vaccine effectiveness data were collected from the epidemiological literature.
Data were collected from mostly healthy adults, typically aged 18--65.
Epidemiological studies of influenza B are less common than those
of influenza A, so studies with ideal experimental design are fewer
in number. In roughly half of the studies used here, there was a minority
of subjects in the range of 13\% to 30\% with conditions such as pregnancy,
cardiovascular disease, diabetes, or body mass index of 40 kg/m$^{2}$
or greater. Interestingly, in all studies, the fraction of subjects
testing positive for influenza B was higher in the subjects without
comorbidity than in the subjects with comorbidity. All data on infections
are from laboratory-confirmed samples, e.g.\ by RT-PCR or the HI
test. Studies focusing on elderly people and children were excluded
to avoid noise caused by the response of an immature immune system
in children or the immunosenescence phenomenon in elder people. Only
studies that used an inactivated vaccine, such as trivalent inactivated
vaccine (TIV), were considered, to decrease bias caused by differences
of vaccine type. The epidemiological data were grouped by season.
For each study, the vaccine effectiveness was calculated from \cite{gupta2006quantifying}
\begin{equation}
E=\frac{u-v}{u}
\label{effectiveness}
\end{equation}
where $u$ is the rate at which unvaccinated people are infected with
influenza, and $v$ is the rate at which vaccinated people are infected
with influenza,
with $u = n_u/N_u$ and $v = n_v/N_v$,
where 
the total number of vaccinated subjects is $N_{v}$, 
the total number of unvaccinated subjects is $N_{u}$, 
the number of influenza B cases among the vaccinated subjects is $n_{v}$,
and 
the number of influenza B cases among the unvaccinated subjects is $n_{u}$. 
 Binomial statistics were assumed
to calculate the error bars for the estimated effectiveness \cite{gupta2006quantifying}:
$\varepsilon=\sqrt{\sigma_{v}^{2}/u^{2}/N_{v}+(v/u^{2})^{2}\sigma_{u}^{2}/N_{u}}$,
where $\sigma_{v}^{2}=v(1-v)$ and $\sigma_{u}^{2}=u(1-u)$. For seasons
that contained $N$ studies, we used the average of the $N$ effectiveness
values, and the standard error was $\varepsilon=\sqrt{\sum_{i}\varepsilon_{i}^{2}/N^{2}}$
where $\varepsilon_{i}$ is the standard error of the $i$th study.

\subsection{Sequence Clustering Analysis}

To gain a geometric understanding of historical influenza B evolution,
we clustered the Influenza Virus Database HA1 sequences from the years
1960 to 2014. The multidimensional scaling method was used to reduce
the sequences from 348 amino acid dimensions to the 3 or 4 dimensions
that best reproduce the Hamming distances between all sequences \cite{Gower}.
Details of this dimensional reduction
 are provided in the Supplemental Information.

\section{Results}

\subsection{Epitope Determination}


The composition of the five epitope regions in the
Victoria and Yamagata models were determined (Table \ref{table1}). 
Note that amino acid site
163 in the Victoria model (B/Brisbane/60/2008) corresponds to a gap
in the Yamagata model (B/Yamanashi/166/1998). Thus, amino acid site number
$i$, $i\ge164$, in the Victoria model corresponds to amino acid site number
$i-1$ in the Yamagata model. In this table, the six amino acid sites that were
added to the epitope by the entropy criterion are shown in red. Additionally,
the eight amino acid sites that were added to the epitope by both the entropy
criterion and RSA are shown in red as well as in
bold style. In epitopes A, B, and D there are two added high-entropy
surface amino acid sites each. In epitope C and E, there is one amino acid site added
to each epitope region. 

In the Victoria model, there are 22 amino acids in epitope A, 25 in
epitope B, 23 in epitope C, 40 in epitope D and 25 in epitope E. In
the Yamagata model, epitope sizes of A, C, D, and E are the same as
in the Victoria model, while epitope B is 24 amino acids due to the
gap. 

\subsection{Lineage Model Chosen}

The main difference between the Victoria and Yamagata lineages
is the gap of amino acid site 163 (Victoria numbering). This gap cannot be
ignored because it is on epitope B, while the nearby amino acid sites 160
to 165 are also on epitope B. We used the methods described in Section
\ref{method:antigenic_distance} to estimate
the vaccine effectiveness
from several
sequence-based measures of antigenic distance. When the vaccine lineage is
Yamagata, amino acid site 163 (Victoria numbering) is eliminated from the
antigenic distance calculation.

\subsection{Vaccine Effectiveness Correlates with Antigenic Distance}

\begin{sidewaysfigure}
\caption{Vaccine effectiveness declines with antigenic distances\label{fig1}}

\begin{centering}
a) \includegraphics[width=3.5in,clip=]{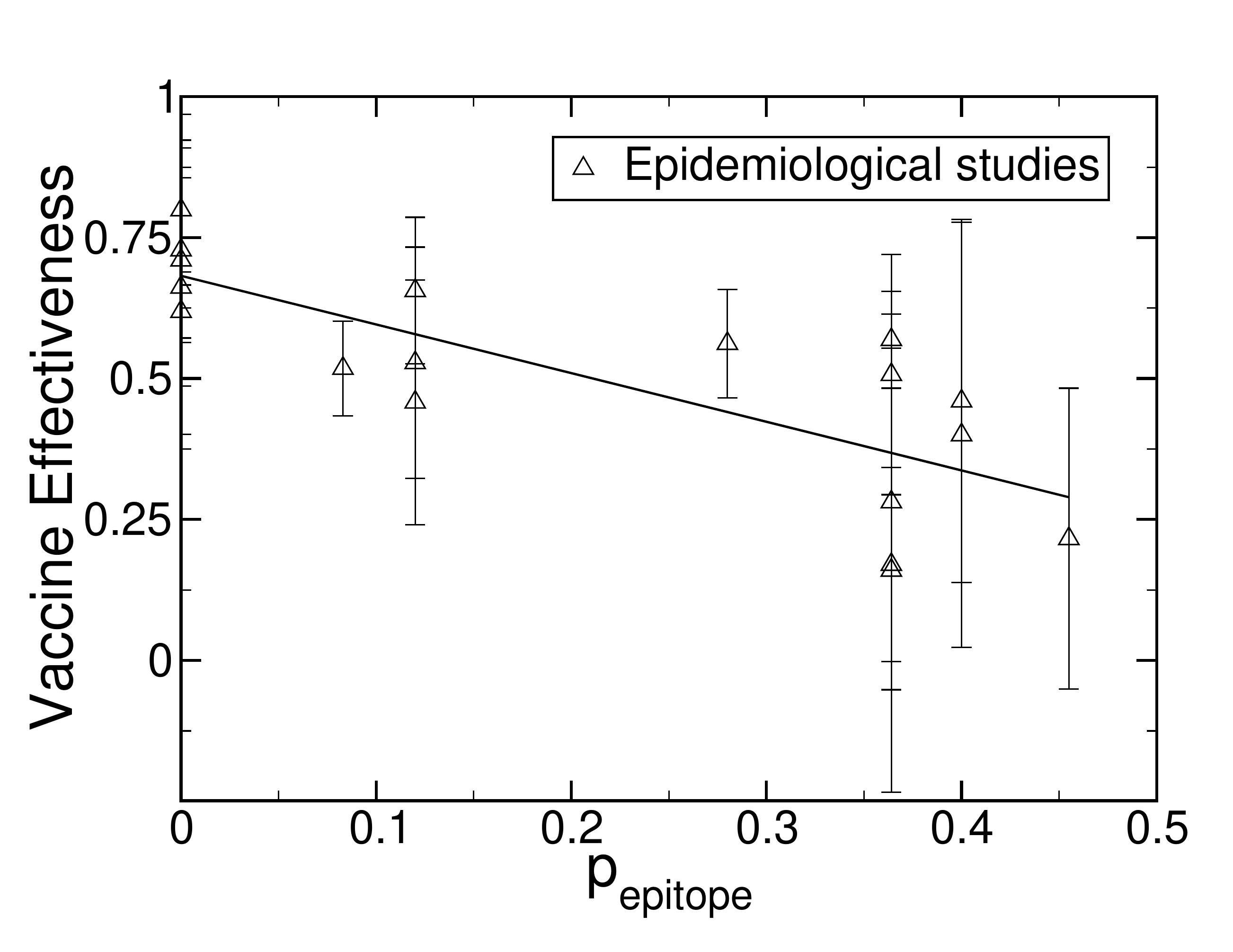}
b) \includegraphics[width=3.5in,clip=]{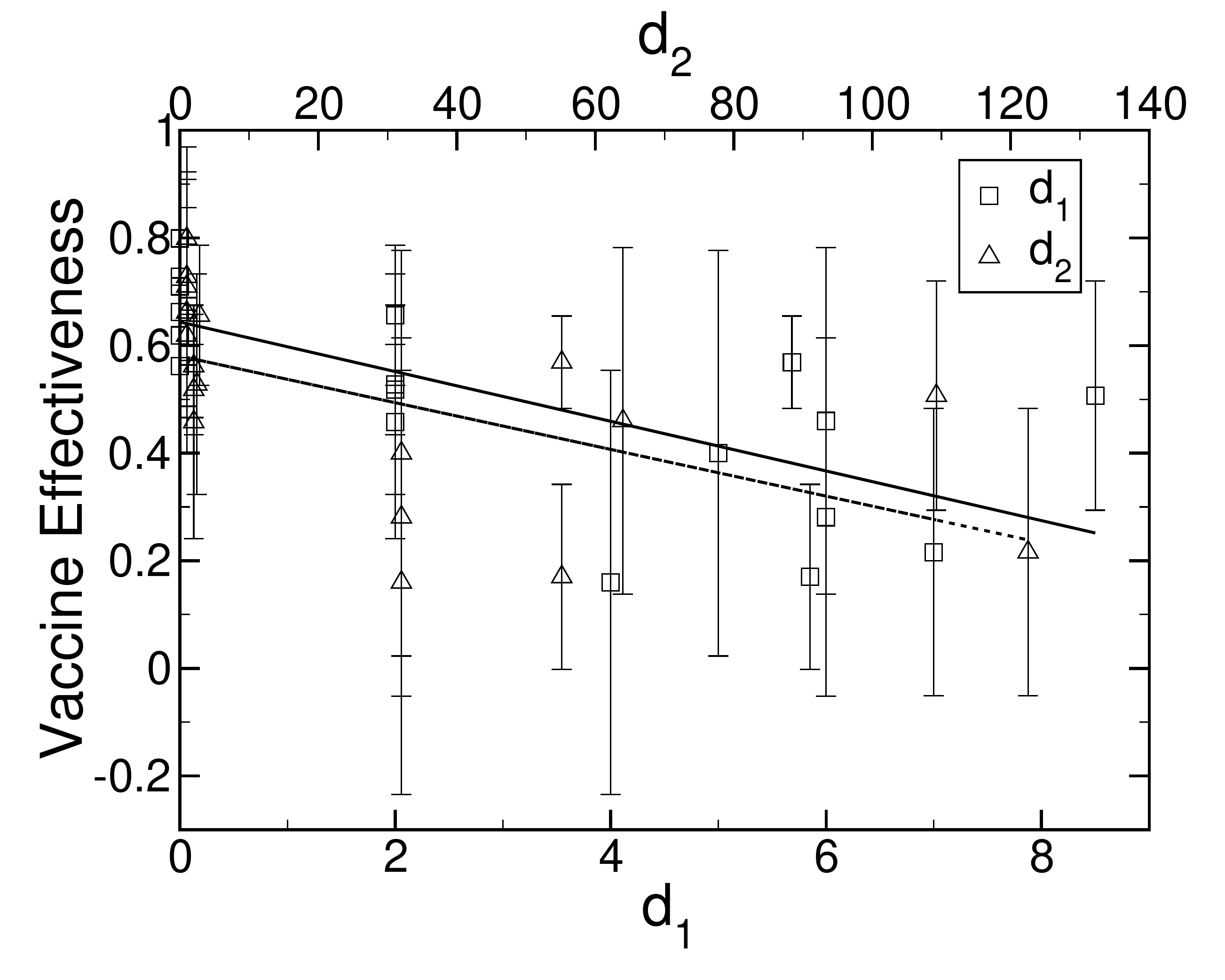}\\
c) \includegraphics[width=2.1in,clip=]{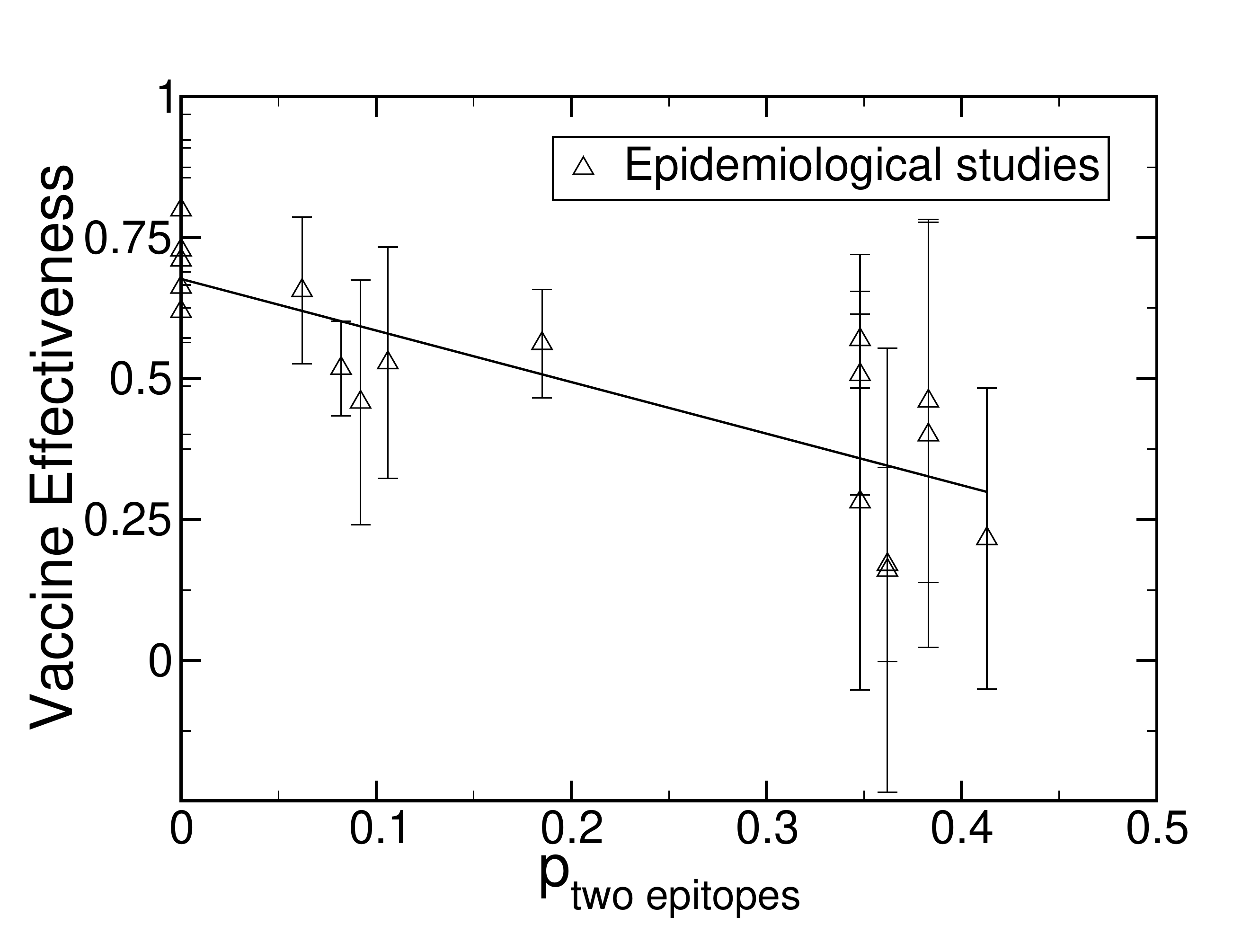}
d) \includegraphics[width=2.1in,clip=]{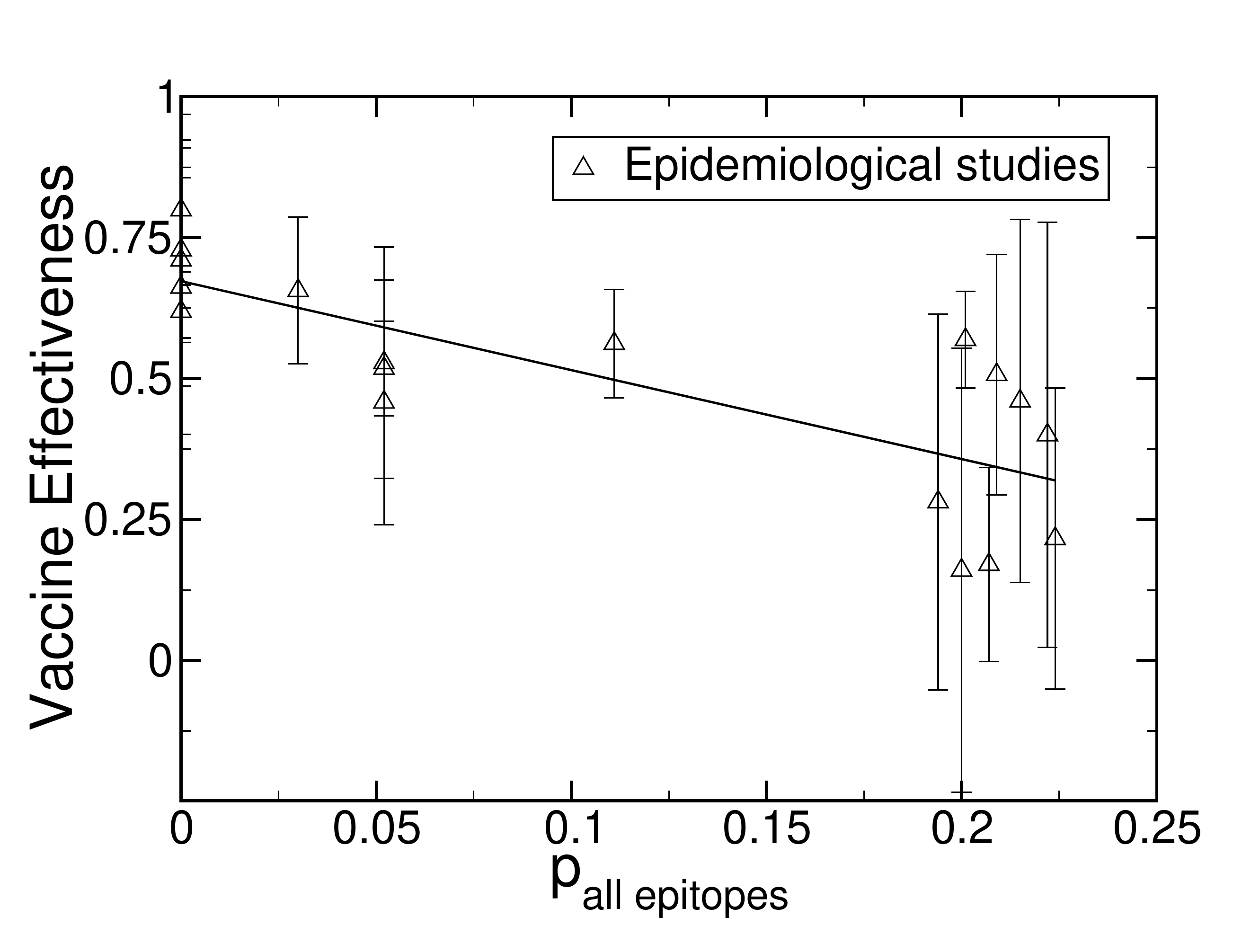}
e) \includegraphics[width=2.1in,clip=]{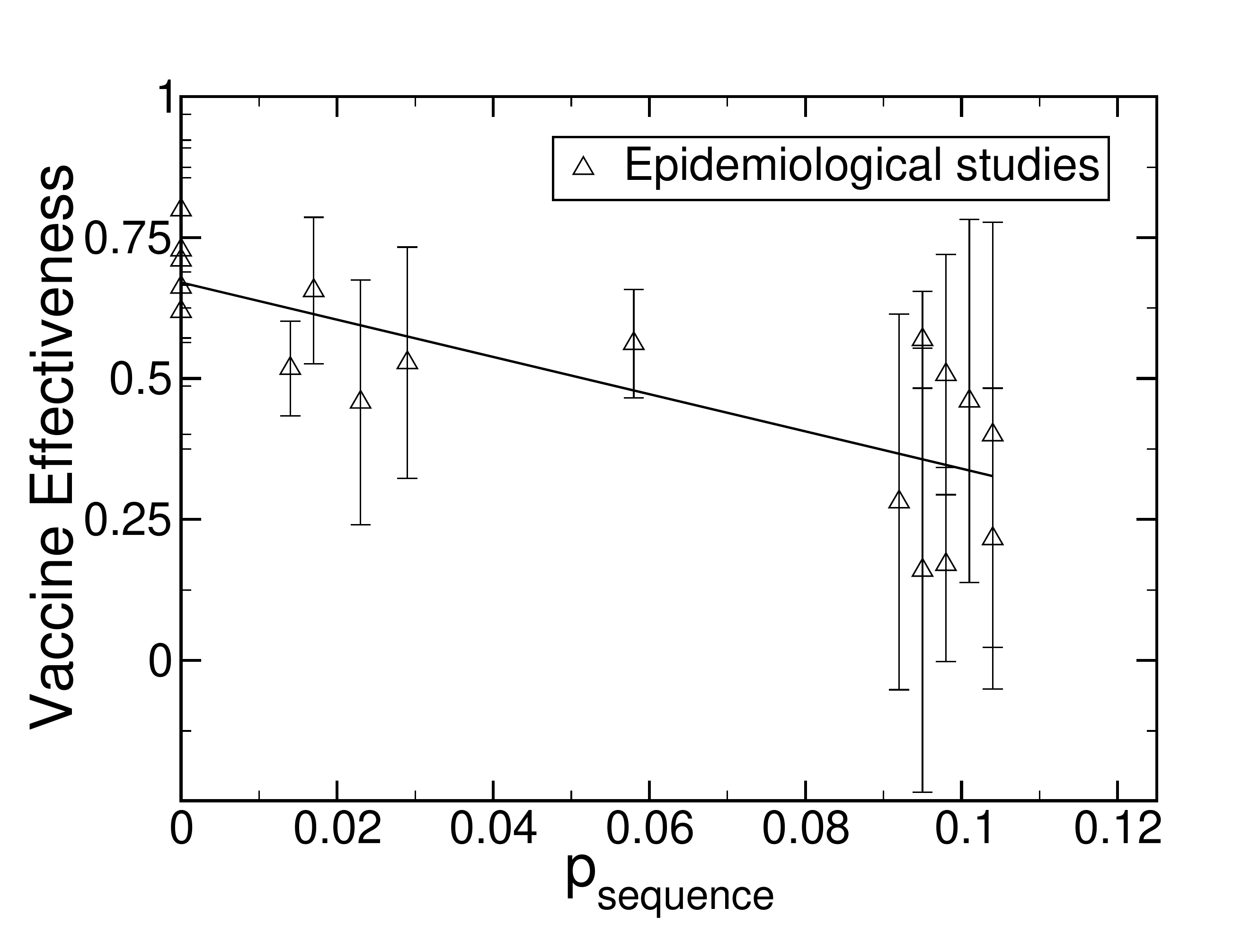}\\
\end{centering}

Measures of antigenic distance.
Correlation of vaccine effectiveness in humans
with a) $p_{{\rm epitope}}$,
b) two standard distances extracted from ferret
animal model data \cite{gupta2006quantifying,pan2011novel},
c) $p_{{\rm two\ epitopes}}$, 
d) $p_{{\rm all\ epitopes}}$, and 
e) $p_{{\rm sequence}}$.
Linear least squares fit of each chart are $p_{{\rm epitope}}$:
$E=-0.864\ p_{{\rm epitope}}+0.6824$,
$R^2 = 0.61$;
$d_1$: 
$E=-0.0462 d_1 + 0.6439$,
$R^2 = 0.48$; 
$d_2$:
$E=-0.0028 d_2 + 0.5805$,
$R^2 = 0.31$; 
$p_{{\rm two\ epitopes}}$:
$E=-0.9151\ p_{{\rm two\ epitopes}}+0.6767$,
$R^2 = 0.65$; 
$p_{{\rm all\ epitopes}}$:
$E=-1.5768\ p_{{\rm all\ epitopes}}+0.6727$,
$R^2 = 0.62$; 
 and
 $p_{{\rm sequence}}$:
$E=-3.3016\ p_{{\rm sequence}}+0.6706$,
$R^2 = 0.61$.
\end{sidewaysfigure}

The influenza B vaccine effectiveness values calculated by Eq.\ \ref{effectiveness}
from epidemiological studies are listed in Table \ref{table2}. This
table also presents the antigenic distances calculated from the epitopes
defined above between the vaccine strain and the dominant circulating strain.
Three additional measures of distance are also listed: $p_{{\rm two~epitopes}}$,
$p_{{\rm all~epitopes}}$, and $p_{{\rm sequence}}$. Figure \ref{fig1}
represents how vaccine effectiveness declines with antigenic distance.
A linear least squares fit shows that $E=-0.864\ p_{{\rm epitope}}+0.6824$. 

From this equation, the antigenic distance at which vaccine effectiveness
declines to zero is $p_{{\rm epitope}}=0.79$. Using the size of each
epitope and the frequency with which each epitope is dominant from
Table \ref{table2}, we calculated the average number of amino acid
substitutions to which this value of $p_{{\rm epitope}}$ corresponds
(see Table S3), 19 for the $p_{{\rm epitope}}$ metric.
The number of substitutions
at which the expected vaccine effectiveness declines to zero according
to the linear least squares fits for the other three measures of antigenic
distance are also displayed in this table.

A further sensitivity analysis was performed, limiting the data to
laboratory genetically confirmed samples, such as by RT-PCR. There were 19
out of 25 data points meeting this requirement. The linear least squares fit
using only those data is
 $E=-0.8252\ p_{{\rm epitope}}+0.684$, and $R^2=0.58$.
This result is quite similar to that of Figure \ref{fig1}a,
 thus demonstrating limited
sensitivity to such a data restriction.
Since RT-PCR was lacking mostly in the early years,  1979-1988, and a relatively
small fraction of the population was
vaccinated in those years, the bias
stemming from HI is expected to be relatively small.

\subsection{Dynamics of Influenza B Evolution}

\begin{sidewaysfigure}
\caption{3D and 4D Dimensional reduction results\label{fig2}}

\begin{centering}
\includegraphics[scale=0.35]{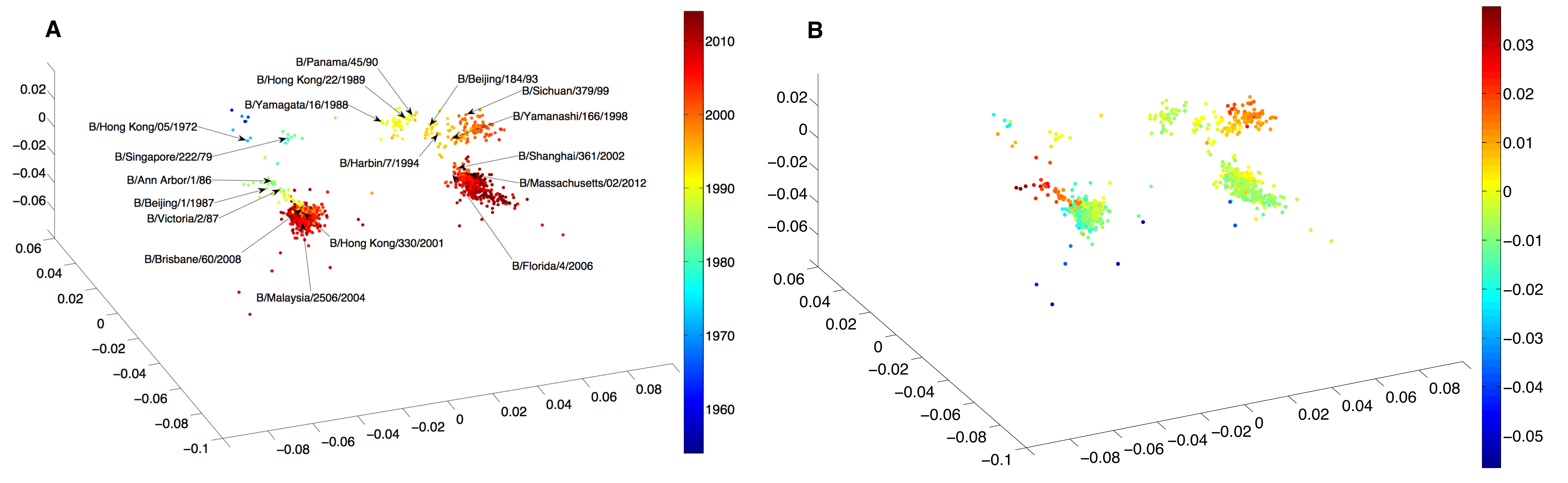}
\par\end{centering}

Dimensional reduction of all influenza B HA1 sequences from Homo sapiens,
obtained from Influenza Virus Database of NCBI \cite{bao2008influenza}.
Data were collected from year 1960 to 2014. Unit of axes are inverse
length of HA1 sequence, 348 amino acids.
a) 3D dimensional reduction. Also
highlighted are 18 representative
influenza B strains. The format of lineage names is \{strain\}/\{sequence-number\}/\{collection-date\}. The collection
years are indicated by the color bar. b) 4D Dimensional reduction.
The fourth dimension in the 4D figure is indicated by the color bar.
\end{sidewaysfigure}

\begin{sidewaysfigure}
\caption{Contour map\label{fig4}}

\begin{centering}
\includegraphics[scale=0.30]{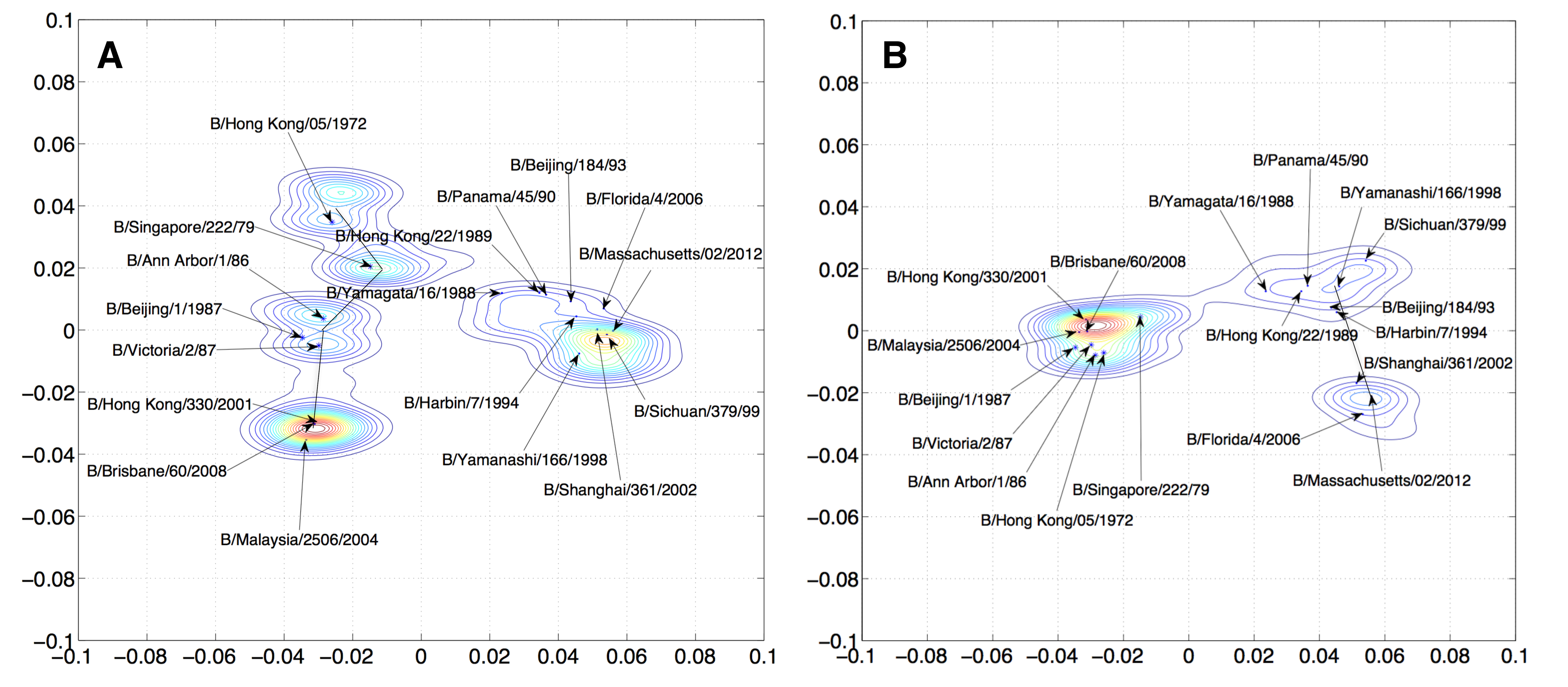}
\par\end{centering}

The a) $x$-$y$ and b) $x$-$z$ projection of contour maps of sequence
clustering. The 18 representative lineages shown in Figure \ref{fig2}
are labeled. Measured distances between the major cluster transitions are shown as solid lines.
\end{sidewaysfigure}

Multidimensional scaling was used to reduce the dimensions to the three
most informative dimensions. Figure \ref{fig2}a shows the evolution
of influenza B from 1960 to 2013. 
Numbers of sequences per year are shown in Table S2.
The left cluster and older points
belong to the Victoria lineage, and the two clusters on the right are
from the Yamagata lineage. This Figure illustrates that the influenza
B strain split into the two Victoria and Yamagata lineages in the
1980s, both of which started to co-circulate from that time. Figure
\ref{fig2}a also suggests that the initial Yamagata lineage from
the 1980s gave way to a distinct daughter lineage around 2000. Note
that the third dimension is necessary to resolve this splitting of
the Yamagata lineage. Analysis of the phylogenetic tree is consistent
with the clustering result shown in Figs. \ref{fig2} and \ref{fig4}. 

Because of the small difference in the eigenvalue associated with the
third dimension and the fourth dimension, we also reduced the dimensions
to a 4D space. Figure \ref{fig2}b shows the evolution, while the color
coding represents the fourth dimension. This Figure shows that
the Yamagata strain jumped out and back 
in the fourth dimension in the late 90s. 

Two contour graphs were obtained from the Gaussian kernel density
estimation. The results are shown in Figure \ref{fig4}. Figure \ref{fig4}a
is the $x$-$y$ axis projection, and Figure \ref{fig4}b is the $x$-$z$
projection. Several representative influenza B sequences are indicated.
From Figure \ref{fig4}, we can measure the distance between peaks,
the evident clusters in density estimation. We measured the average
distance between the major cluster transitions, shown as the dark
lines in Figure \ref{fig4}. The average number of substitutions is
approximately 12. We also measured the full width at half maximum
for the major clusters. We find a width of approximately 6 substitutions.
The corresponding peak-peak distance and cluster width for H3N2 are
6 and 3 amino acids, respectively. \cite{clustering}

\subsection{Distribution of Epitopes and High-Entropy Residues}

\begin{figure}
\caption{Epitopes and additional high-entropy amino acid sites on the HA1 3D structure\label{fig5}}

\begin{centering}
\includegraphics[scale=0.30]{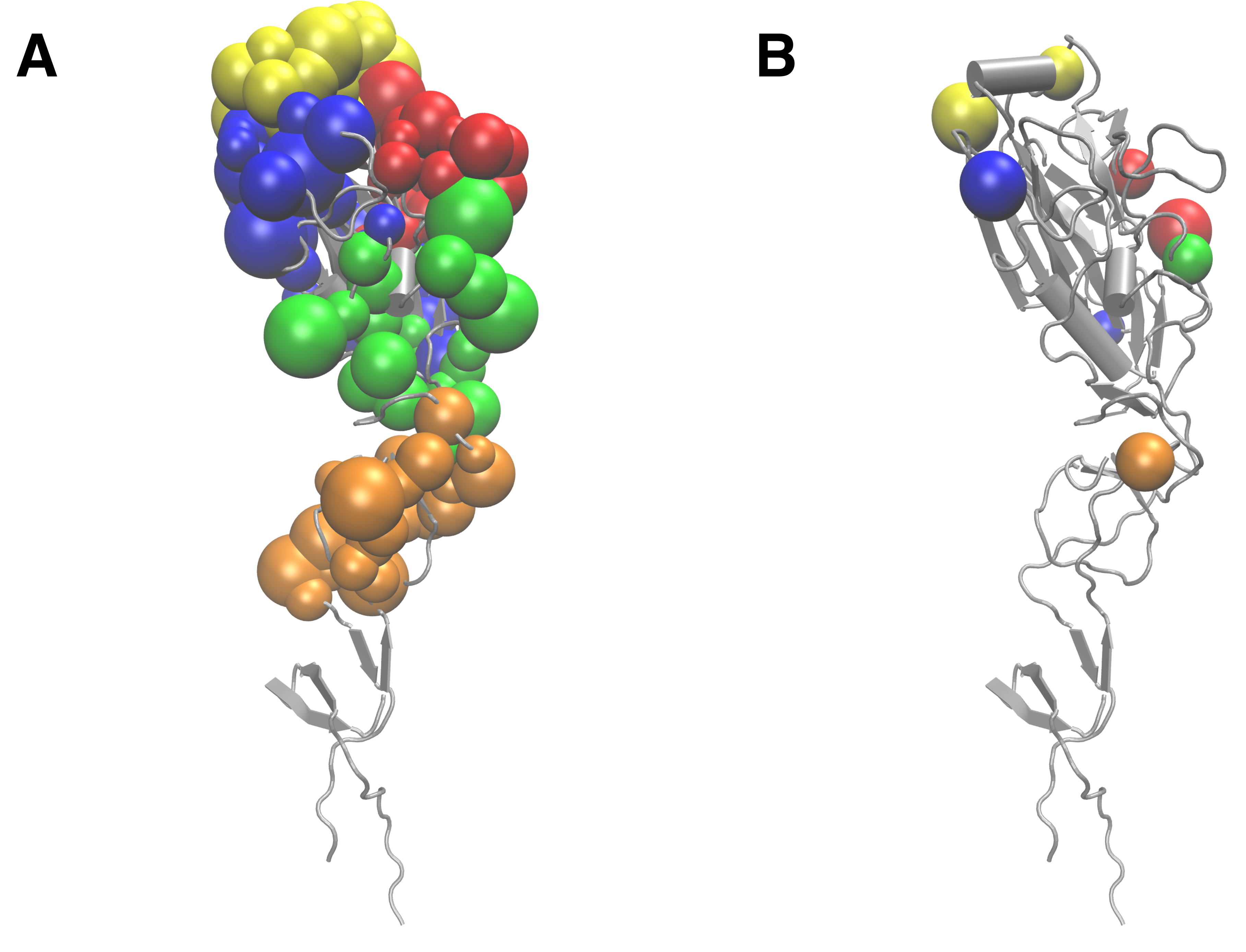}
\par\end{centering}

a: Three dimensional structure of influenza B, HA1 domain (Victoria
lineage, model 4FQM) with five epitopes. Epitopes are shown in
space filling representation
with different colors corresponding to the five epitopes:
epitope A (red), epitope B (yellow), epitope C (orange),  epitope D (blue),
epitope E (green), and non-epitope region (silver).
b: Three dimensional structure of influenza B, HA1 domain (Victoria lineage)
with the 8 high-entropy amino acid sites newly added to the epitope region:
amino acid site 122 and 126 to epitope A, 166 and 209 to epitope B, 40 to
epitope C, 233 and 255 to epitope D, and 73 to epitope E. Here the
Victoria numbering is used.

\end{figure}

Figure \ref{fig5}a reveals the HA1 domain of influenza B (PDB ID:4FQM).
This structure is for the Victoria-like lineage. The structure 
for the Yamagata-like lineage is visually indistinguishable. The five
epitopes are represented as five space-filling regions with distinct
colors. The epitope amino acid sites are on the surface of HA1. Figure \ref{fig5}b
shows the 8 amino acid sites added to the epitope region because of both high
seasonal-average entropy and large RSA. These additional amino acid sites are
color coded according to the five epitopes, as in Figure \ref{fig5}a.
Figure S2
shows the average seasonal entropy of each amino
acid in sequence of the HA1 domain. Each amino acid site is color coded according
to the five epitopes.

\subsection{Dispersed Dominant Epitope in Each Season}

The dominant epitope varies by season. Table S4
describes
the fractional change of each epitope in each season, sorted according
to magnitude. By definition, the first column is $p_{{\rm epitope}}$.
Epitope B was most frequently dominant (7 seasons), although A was
also quite frequently dominant (6 seasons). Epitope C always had the
least fractional change. The next most dominant epitope also varies by
seasons: it could be A, B, D, or E. Note that the dominant and next
most dominant epitopes could have similar fractional
changes in amino acids,
such as in 2012-2013 (b) and 2011-2012 (b).

\subsection{Evolution of Influenza B}

In this study, we used dimensional reduction to characterize the evolution
of influenza B. Phylogenetic trees are consistent with these results.
We constructed a neighbor joining tree and a maximum likelihood tree based
on the same sequence set for which we performed dimensional
reduction. In these trees
the Yamagata lineages and Victoria lineages are divided into two branches
in the tree, originating from sequences obtained in the 1980s. These
trees are very dense, however,
and are hard to use to distinguish
distances between clusters. Thus, we provide a simplified phylogenetic
tree in Figure S1.
We can see the Yamagata and Victoria branches
in Figure S1.
The distances among each cluster and sub-cluster
in the dimensionally reduced space are clearer than what the trees can provide.
The dimensionally reduced space is an optimal representation of relative
distances, unlike the phylogenetic trees that are constrained to a
tree topology.

We further analyzed the distance among peaks in our Gaussian kernel
density estimation. Based on Figure \ref{fig4}, the average distance
between major peaks in the same lineage is approximately 12 amino acid sites.
In this analysis, peaks in Victoria lineage and peaks in Yamagata
lineage were measured separately. Influenza B strains in the early
years belonged to the Victoria lineage and then separated to two lineages
in the 1980s.
There is a small transitional peak in between these two lineage clusters
The distance
between these two lineages is approximately twice the average distance
of 12 amino acid sites.

\section{Discussion}

\subsection{Antigenic Distance at which Vaccine
Effectiveness is Zero}

The number of substitutions at which 
vaccine effectiveness is predicted to decline
to zero is shown
in Table S3.
For the $p_{epitope}$ estimate,
vaccine effectiveness decays to zero at roughly 
19 substitutions.
This result of the
$p_{epitope}$ estimate
is the closest to the average distance
observed between the evolving strain clusters in
the dimensional reduction analysis, 12 amino acid sites. 
Thus, we chose $p_{epitope}$ as our metric for antigenic distance. 
In a previous
study of H3 in influenza A \cite{clustering}, the average distance
between consecutive H3 clusters was 6 amino acid sites, and the average substitution
number when vaccine effectiveness decreases to zero was approximately
4. It appears influenza clusters emerge at just enough distance
to evade prior immunity, whether induced by prior infection or
vaccination in the human population.

\subsection{Lower Selective Pressure Compared to Influenza A}

Influenza A has high selection pressure, high mutant rate, and distinct
dominant epitopes in each season \cite{deem2009epitope,pan2011novel,xi2015H3}.
Influenza A is generally more common than influenza B. Influenza
B can, however, predominate in certain
regions in each influenza season \cite{WHOeachyearIB,hay2001evolution}.
Since influenza A is widely predominant, it may be under more selective
pressure to evolve. The influenza A vaccine commonly contains one lineage
from A/H1N1 and one from A/H3N2 \cite{WHOeachyearIB}. These vaccines
cause antibodies bind to the virus in a specific manner and stimulate the evolution of
the dominant epitope. Vaccine and prior immune history are likely the reasons
for the distinct dominant epitopes in influenza A. 

Unlike influenza A, the dominant epitopes of influenza B are not 
clear, and the high-entropy amino acid sites are dispersed. These results
indicate that selective pressure of influenza B may be more dispersed
on the protein surface. For certain regions of the influenza B epitopes,
the selective pressure may relatively be low, and antibody binding may
be in a more random pattern. These results explain why in some seasons,
the second most dominant epitope often has a similar fraction of amino
acid substitutions as the dominant epitope.

\section{Conclusion}
In this paper we have considered how influenza B vaccine
effectiveness depends on antigenic distance. 
We have defined a measure of antigenic distance
for influenza B,
$p_{{\rm epitope}}$.
This measure is defined as the fraction of 
amino acid site substitutions between the vaccine and circulating
virus lineage in the dominant epitope region
of HA1 \cite{gupta2006quantifying}.
This measure correlates well with influenza B vaccine
effectiveness in humans. We find that
new, emergent strains of influenza B tend
to occur at large values of $p_{\rm epitope}$, for which
immune recognition will be minimal.  In other words,
influenza B evolves to escape immune recognition due
to prior infection or vaccination in the human population.

This measure of antigenic distance
provides a novel tool for prediction 
of vaccine effectiveness and
may be used in the context of annual influenza vaccine design.
The dimensional reduction technique illustrated here can
be used to identify incipient dominant strains.  In
conjunction with other available data, such as ferret animal
data, the $p_{\rm epitope}$ measure can be used to estimate whether
a new vaccine strain will be required, by measuring the antigenic
distance between the existing vaccine strain and the incipient strain.
The $p_{\rm epitope}$ measure applies to both the Victoria
and Yamagata lineages.
The $p_{\rm epitope}$ measure of antigenic distance may be used,
again in conjunction with other available data,
to select which vaccine strain in each lineage would be predicted to be most
protective against the distribution of predicted or
observed circulating viral strains.
It is also possible to perform this prediction on a geographically
localized scale, predicting vaccines optimal for different parts
of the world.
The $p_{\rm epitope}$ measure of antigenic distance may 
 also be used
to select among egg-viable, ``like'' strains for those predicted
to be most antigenically similar to the desired vaccine strain.
These capabilities complement existing analysis of sequence
and animal model datasets.

\section*{Acknowledgments}

\hbox{}\hspace{-\parindent}\textbf{Competing Financial Interests}
Authors declare no competing financial interests.

\hbox{}\hspace{-\parindent}\textbf{Correspondence} Correspondence
should be addressed to MWD (mwdeem@rice.edu). 

\clearpage{}

\begin{table}[h]
\caption{Residues in epitope region of influenza B \label{table1} }
\begin{tabular}{lll}
\hline 
 & Epitope  & Residue \tabularnewline
\hline 
 & A  & 121 \textbf{\textcolor{red}{122}} 123 125 \textbf{\textcolor{red}{126}}
134 135 136 137 139 141 142 144 \tabularnewline
 &  & 146 147\textbf{\textcolor{red}{{} }}\textcolor{red}{148} 149 150 151
155 157 177 \tabularnewline
 & B  & \textcolor{red}{127} 129 133 160 161 162 163 164 165 \textbf{\textcolor{red}{166}}
168 \textcolor{red}{172} 174\tabularnewline
 &  & 196 197 198 199 200 202 203 204 206 207 208 \textbf{\textcolor{red}{209}}\tabularnewline
Victoria model & C  & 34 35 36 37 38 39 \textbf{\textcolor{red}{40}} 289 291 292 293 294
309 315\tabularnewline
 &  & 317 318 320 321 323 324 325 326 327 \tabularnewline
 & D  & 93 101 102 116 120 176 179 \textcolor{red}{180} 182 183 184 185 186 \tabularnewline
 &  & 187 188 190 212 214 218 219 220 223 224 225 226 \tabularnewline
 &  & 227 228 229 230 \textbf{\textcolor{red}{233}} 242 243 244 245 246
254 \textbf{\textcolor{red}{255}}\tabularnewline
 &  & 256 257 258 \tabularnewline
 & E  & 42 44 \textcolor{red}{48} 56 \textcolor{red}{58} 59 63 71 \textbf{\textcolor{red}{73}}
75 77 78 79 80 83 84 85\tabularnewline
 &  & 88 89 91 108 273 276 277 280 \tabularnewline
\hline 
 & A  & 121 \textbf{\textcolor{red}{122}} 123 125 \textbf{\textcolor{red}{126}}
134 135 136 137 139 141 142 144\tabularnewline
 &  & 146 147 \textcolor{red}{148} 149 150 151 155 157 176 \tabularnewline
 & B  & \textcolor{red}{127} 129 133 160 161 162 163 164 \textbf{\textcolor{red}{165}}
167 \textcolor{red}{171} 173 195 \tabularnewline
 &  & 196 197 198 199 201 202 203 205 206 207 \textbf{\textcolor{red}{208}}\tabularnewline
Yamagata model  & C  & 34 35 36 37 38 39 \textbf{\textcolor{red}{40}} 288 290 291 292 293
308 314\tabularnewline
 &  & 316 317 319 320 322 323 324 325 326 \tabularnewline
 & D  & 93 101 102 116 120 175 178 \textcolor{red}{179} 181 182 183 184 185\tabularnewline
 &  & 186 187 189 211 213 217 218 219 222 223 224 \tabularnewline
 &  & 225 226 227 228 229 \textbf{\textcolor{red}{232}} 241 242 243 244
245 253 \textbf{\textcolor{red}{254}}\tabularnewline
 &  & 255 256 257 \tabularnewline
 & E  & 42 44 \textcolor{red}{48} 56 \textcolor{red}{58} 59 63 71 \textbf{\textcolor{red}{73}}
75 77 78 79 80 83 84 85\tabularnewline
 &  & 88 89 91 108 272 275 276 279 \tabularnewline
\hline 
\end{tabular}

Residues in epitopes A, B, C, D, and E of influenza B. The results
for both the Victoria model (B/Brisbane/60/2008 numbering) and the
Yamagata model (B/Yamanashi/166/1998 numbering) are shown. 
The six amino acid sites that were added to the epitope by the entropy criterion 
are shown in red.  
The eight amino acid sites that were added to the epitope by both the entropy
criterion and relative accessible surface area (RSA) are shown in red bold.
\end{table}

\clearpage

\begin{sidewaystable}
 \caption{Vaccine strains, dominant circulating strains, vaccine effectiveness,
and antigenic distances \label{table2} }
{\tiny{}}%
\begin{tabular}{llllllllllll}
\hline 
{\tiny{}Season } & {\tiny{}Vaccine } & {\tiny{}Dominant } & {\tiny{}VE } & {\tiny{}${\varepsilon_{VE}}$ } & {\tiny{}Dominant } & {\tiny{}$p_{{\rm epitope}}$} & {\tiny{}$p_{{\rm two\ epitopes}}$ } & {\tiny{}$p_{{\rm all\ epitopes}}$ } & {\tiny{}$p_{{\rm sequence}}$ } & {\tiny{}$d_1$} &{\tiny{}$d_2$}\tabularnewline
 & {\tiny{}Strain } & {\tiny{}Strain } &  &  & {\tiny{}Epitope } &  &  &  &  & \tabularnewline
\hline 
{\tiny{}2012-2013 (a) } & {\tiny{}B/Wisconsin/1/2010[Y] (AET22022) } & {\tiny{}B/Wisconsin/1/2010[Y] (AET22022) } & {\tiny{}0.619 \cite{centers2013interim,mclean2015influenza}} & {\tiny{}0.047 } & {\tiny{}- } & {\tiny{}0 } & {\tiny{}0 } & {\tiny{}0 } & {\tiny{}0 } &{\tiny{}0 } &{\tiny{}1 } \tabularnewline
{\tiny{}2012-2013 (b) } & {\tiny{}B/Wisconsin/1/2010[Y] (AET22022) } & {\tiny{}B/Massachusetts/2/2012[Y] (AGL06036) } & {\tiny{}0.518 \cite{mcmenamin2013effectiveness,kissling2014influenza}} & {\tiny{}0.084 } & {\tiny{}B } & {\tiny{}0.083 } & {\tiny{}0.082} & {\tiny{}0.052} & {\tiny{}0.014 } &{\tiny{}2{\cite{barr2014recommendations}} } &{\tiny{}2 } \tabularnewline
{\tiny{}2012-2013 (c) } & {\tiny{}B/Wisconsin/1/2010[Y] (AET22022) } & {\tiny{}B/Brisbane/60/2008[V] (ACN29380) } & {\tiny{}0.569 \cite{mclean2015influenza} } & {\tiny{}0.086 } & {\tiny{}A } & {\tiny{}0.364} & {\tiny{}0.348} & {\tiny{}0.201} & {\tiny{}0.095 } & {\tiny{}5.68{\cite{ran2015domestic,lin2015optimisation}} } & {\tiny{}55.16 }\tabularnewline
{\tiny{}2011-2012 (a) } & {\tiny{}B/Brisbane/60/2008[V] (ACN29380) } & {\tiny{}B/Bangladesh/3333/2007[Y] (AFH58304) } & {\tiny{}0.460 \cite{castilla2013decline}} & {\tiny{}0.322 } & {\tiny{}B } & {\tiny{}0.400} & {\tiny{}0.383} & {\tiny{}0.215} & {\tiny{}0.101 } &{\tiny{}6{\cite{barr2010epidemiological,flu2011Switzerland}} }&{\tiny{}64 } \tabularnewline
{\tiny{}2011-2012 (b) } & {\tiny{}B/Brisbane/60/2008[V] (ACN29380) } & {\tiny{}B/Wisconsin/01/2010[Y] (AET22022) } & {\tiny{}0.170 \cite{skowronski2014influenza}} & {\tiny{}0.172 } & {\tiny{}A } & {\tiny{}0.364} & {\tiny{}0.362} & {\tiny{}0.207} & {\tiny{}0.098 } & {\tiny{}5.85{\cite{ran2015domestic,lin2015optimisation}} }&{\tiny{}55.16 } \tabularnewline
{\tiny{}2011-2012 (c) } & {\tiny{}B/Brisbane/60/2008[V] (ACN29380) } & {\tiny{}B/Brisbane/60/2008[V] (ACN29380) } & {\tiny{}0.799 \cite{pebody2013vaccine,skowronski2014influenza}} & {\tiny{}0.110 } & {\tiny{}-} & {\tiny{}0 } & {\tiny{}0 } & {\tiny{}0 } & {\tiny{}0 } & {\tiny{}0 }&{\tiny{}1 } \tabularnewline
{\tiny{}2011 } & {\tiny{}B/Brisbane/60/2008[V] (ACN29380) } & {\tiny{}B/Brisbane/60/2008[V] (ACN29380) } & {\tiny{}0.728 \cite{levy2014influenza,kelly2013moderate}} & {\tiny{}0.241 } & {\tiny{}-} & {\tiny{}0 } & {\tiny{}0 } & {\tiny{}0 } & {\tiny{}0 } & {\tiny{}0 } & {\tiny{}1 }  \tabularnewline
{\tiny{}2010 } & {\tiny{}B/Brisbane/60/2008[V] (ACN29380) } & {\tiny{}B/Brisbane/60/2008[V] (ACN29380) } & {\tiny{}0.710 \cite{levy2014influenza}} & {\tiny{}0.146 } & {\tiny{}- } & {\tiny{}0 } & {\tiny{}0 } & {\tiny{}0 } & {\tiny{}0 } & {\tiny{}0 } & {\tiny{}1 }  \tabularnewline
{\tiny{}2008-2009 } & {\tiny{}B/Florida/04/2006[Y] (ACA33493) } & {\tiny{}B/Brisbane/60/2008[V] (ACN29380) } & {\tiny{}0.507 \cite{barrett2011efficacy}} & {\tiny{}0.213 } & {\tiny{}A } & {\tiny{}0.364} & {\tiny{}0.348} & {\tiny{}0.209} & {\tiny{}0.098 } & {\tiny{}8.50{\cite{barr2010epidemiological,flu2011Switzerland}}} &{\tiny{}109.26 } \tabularnewline
{\tiny{}2008} & {\tiny{}B/Florida/04/2006[Y] (ACA33493) } & {\tiny{}B/Florida/04/2006[Y] (ACA33493) } & {\tiny{}0.662  \cite{fielding2011estimation,kelly2013moderate}} & {\tiny{}0.261 } & {\tiny{}-} & {\tiny{}0 } & {\tiny{}0 } & {\tiny{}0 } & {\tiny{}0 } &{\tiny{}0} & {\tiny{}1}  \tabularnewline
{\tiny{}2007-2008} & {\tiny{}B/Malaysia/2506/2004[V] (ACF54235) } & {\tiny{}B/Florida/04/2006[Y] (ACA33493) } & {\tiny{}0.400  \cite{frey2010clinical,monto2009comparative,treanor2011protective}} & {\tiny{}0.377 } & {\tiny{}B } & {\tiny{}0.400 } & {\tiny{}0.383 } & {\tiny{}0.222 } & {\tiny{}0.104{\cite{WHO2008}} } &{\tiny{}5 } & {\tiny{}32 } \tabularnewline
{\tiny{}2006-2007} & {\tiny{}B/Malaysia/2506/2004[V] (ACF54235) } & {\tiny{}B/Shanghai/361/2002[Y] (CAH04474) } & {\tiny{}0.160  \cite{skowronski2009component}} & {\tiny{}0.394 } & {\tiny{}A } & {\tiny{}0.364} & {\tiny{}0.362} & {\tiny{}0.200} & {\tiny{}0.095 } & {\tiny{}4{\cite{WHO2008}} } & {\tiny{}32 } \tabularnewline
{\tiny{}2005-2006 (a) } & {\tiny{}B/Shanghai/361/2002[Y] (CAH04474) } & {\tiny{}B/Malaysia/2506/2004[V] (ACF54235) } & {\tiny{}0.281 \cite{skowronski2007estimating}} & {\tiny{}0.333 } & {\tiny{}A } & {\tiny{}0.364 } & {\tiny{}0.348 } & {\tiny{}0.194 } & {\tiny{}0.092 } & {\tiny{}6{\cite{WHO2008}} }&{\tiny{}32 } \tabularnewline
{\tiny{}2005-2006 (b) } & {\tiny{}B/Jiangsu/10/2003[Y] (ACF54191) } & {\tiny{}B/Hong Kong/330/2001[V] (AAT69436) } & {\tiny{}0.216 \cite{beran2009challenge}} & {\tiny{}0.267 } & {\tiny{}A } & {\tiny{}0.455} & {\tiny{}0.413} & {\tiny{}0.224} & {\tiny{}0.104 } &{\tiny{}7{\cite{RecommendWHO2006,WHO2004}} } &{\tiny{}122.51 } \tabularnewline
{\tiny{}1987-1988} & {\tiny{}B/Ann Arbor/1/86[V] (AAB16995) } & {\tiny{}B/Victoria/2/87[V] (P22092) } & {\tiny{}0.458  \cite{keitel1997efficacy}} & {\tiny{}0.217 } & {\tiny{}B } & {\tiny{}0.120} & {\tiny{}0.092} & {\tiny{}0.052} & {\tiny{}0.023 } & {\tiny{}2{\cite{centers1988update}} }&{\tiny{}2 } \tabularnewline
{\tiny{}1985-1986 } & {\tiny{}B/USSR/100/83[V] (P09766) } & {\tiny{}B/Ann Arbor/1/86[V] (ABF21280) } & {\tiny{}0.562 \cite{keitel1997efficacy}} & {\tiny{}0.096 } & {\tiny{}B } & {\tiny{}0.280} & {\tiny{}0.185} & {\tiny{}0.111} & {\tiny{}0.058 } & {\tiny{}0{\cite{centers1988update}} }&{\tiny{}2 } \tabularnewline
{\tiny{}1983-1984 } & {\tiny{}B/Singapore/222/79[V] (P03463) } & {\tiny{}B/USSR/100/83[V] (P09766) } & {\tiny{}0.656 \cite{keitel1997efficacy}} & {\tiny{}0.130 } & {\tiny{}B } & {\tiny{}0.120} & {\tiny{}0.062} & {\tiny{}0.030} & {\tiny{}0.017 } & {\tiny{}2{\cite{hay2001evolution}} }&{\tiny{}2.83 } \tabularnewline
{\tiny{}1979-1980 } & {\tiny{}B/Hong Kong/5/72[V] (ABF21280) } & {\tiny{}B/Singapore/222/79[V] (P03463) } & {\tiny{}0.528 \cite{foy1981ussr}} & {\tiny{}0.205 } & {\tiny{}B } & {\tiny{}0.120 } & {\tiny{}0.106 } & {\tiny{} 0.052} & {\tiny{} 0.029} &{\tiny{}2{\cite{centers1980IB}} }& {\tiny{}2.45 } \tabularnewline
\end{tabular}{\tiny \par}

Identities and accession numbers of vaccine strain and dominant circulating
strain, with corresponding experimental vaccine effectiveness and
standard error. The values of $p_{{\rm epitope}}$, $p_{{\rm two\ epitopes}}$,
$p_{{\rm all\ epitopes}}$, $p_{{\rm sequence}}$, and ferret
HI $d_1$ and $d_2$ data are listed.
Whether the strain is Victoria [V] or Yamagata [Y] is indicated.
 Vaccine effectiveness
is taken from epidemiological studies cited for each influenza season,
and antigenic distances are predicted by the $p_{{\rm epitope}}$
method. Most of these data are obtained from the Northern Hemisphere
(seasons in ``year1-year2'' format). Three seasons from Southern
Hemisphere (2011, 2010, 2008) are also listed.
\end{sidewaystable}

\clearpage

\clearpage

\section*{References}
\bibliography{IB_reference}

\end{document}



\begin{frontmatter}

\title{Supplemental Information for\\
Prediction of Influenza B Vaccine Effectiveness from Sequence Data}

\author[sspb]{Yidan Pan}
\author[bioe,sspb,phys]{Michael W.\ Deem\corref{mycorrespondingauthor}}
\cortext[mycorrespondingauthor]{Corresponding author}
\ead{mwdeem@rice.edu}

\address[sspb]{Systems, Synthetic, and Physical Biology,
Rice University 6100 Main St, Houston, TX  77005}
\address[bioe]{Department of Bioengineering,
Rice University 6100 Main St, Houston, TX  77005}
\address[phys]{Department of Physics \& Astronomy,
Rice University 6100 Main St, Houston, TX  77005}

\end{frontmatter}

\section{Methods}
\subsection{Epitope Mapping}

\label{method:epitope} For influenza A/H3N2, it has been determined
that there are 5 epitopes on the HA1 protein that contribute significantly
to the human immune response to the virus \cite{wilson1990structural}.
These epitopes have been identified by crystallography \cite{wilson1981structure,Epitope2}
and further improved by sequence analysis \cite{ISD,gupta2006quantifying}.
Epitopes for the H1N1 strain of influenza A have been determined by
mapping from these H3N2 epitopes \cite{deem2009epitope,pan2011novel}.
The average root mean square deviation of the mapped epitope region
between influenza A/H3N2 and influenza B Victoria lineage is 3.1517
\AA , and the value between influenza A/H3N2 and influenza B Yamagata
lineage is 3.0931 \AA.

Amino acid sites in epitopes are under selective pressure, which stimulates
their evolution to avoid recognition. This process increases the entropy
value of the amino acid sites in epitope regions. The Shannon entropy
of these amino acid sites per season
was calculated using worldwide HA1 sequences collected
during every Northern Hemisphere influenza season (September to March)
from 1994-1995 season to 2013-2014 season, and then averaged over
all seasons. 
For each epitope amino acid site of influenza A mapping to multiple amino acid sites in influenza
B, the amino acid site with the highest average seasonal Shannon entropy was
included in the epitope \cite{deem2009epitope}. Additionally, those
amino acid sites with an average seasonal Shannon entropy greater than 0.1
were included. This threshold of information entropy is determined
adaptively. Listing amino acid sites in descending order of entropy, 0.1 is
a boundary for non-random entropy values.

We used this threshold to extend the epitope regions. A total of
15 amino acid sites outside the mapped epitope region have an average seasonal
Shannon entropy greater than 0.1. We calculated the relative accessible surface area (RSA) of them based on protein structure 4FQM
using ASA view \cite{shander2004asa}. We included 8 amino acid sites with
RSA greater than 0.5, which were considered amino acid sites on the surface.
These 8 amino acid sites were assigned to epitopes by calculating the distance
of alpha carbons between these amino acid sites and the amino acid sites in epitope region
respectively, assigning amino acid sites to the closet
epitope. For the Victoria model, amino acid site 126 belongs to epitope A. 
However, amino acid site 126 was difficult to assign unambiguously in the Yamagata model.
The minimum distance with a amino acid site in epitope B of the 4M44 Yamagata model was 3.794\AA , while the minimum distance between amino acid site 126 and a amino acid site in epitope
A is 3.808 \AA . Comparing the ten closest epitopic amino acid sites to amino acid site 126
in 4M44, we found that it is closer to epitope A on average. 
Therefore, we determined that amino acid site 126 should be included in epitope A for the Yamagata model.

\subsection{Sequence Clustering Analysis}
The relative importance of each dimension in the
multidimensional scaling is proportional to the eigenvalue
associated with each reduced dimension: 0.05562, 0.02411, 0.00692,
and 0.00640.

In the calculation each sequence was weighted by the inverse of the
number of samples collected per year, to give each year roughly
equal weight in the dimensional reduction procedure. Additionally
Gaussian kernel density estimation was used to estimate the probability
density of sequences in the reduced space identified by the weighted
multidimensional scaling. The standard deviation of the Gaussian weight
was $\sigma_{x}$=0.03351265, $\sigma_{y}$=0.01171378, and $\sigma_{z}$=0.01177186.

The reconstructed probability density of the viruses in the reduced
$(x,y,z)$ space is $P(x,y,z)\propto{\sum_{i}\exp[-\frac{(x-x_{i})^{2}}{2\sigma_{x}^{2}}-\frac{(y-y_{i})^{2}}{2\sigma_{y}^{2}}-\frac{(z-z_{i})^{2}}{2\sigma_{z}^{2}}]}$.
The peaks in this density estimate correspond to clusters of influenza
B sequences. The fourth dimension in the 4D figure is indicated by
the color bar.

The phylogenetic tree of these sequences was constructed using MEGA
6.06 \cite{tamura2013mega6}. A neighbor joining tree was first constructed.
A maximum likelihood tree was then constructed, whose structure confirmed
that of the neighbor joining tree. The maximum likelihood tree was
divided into three groups: Two Yamagata-like lineages and one Victoria-like
lineage.

\section{Discussion}
\subsection{Dispersed High Entropy Amino Acid Sites}

The distribution of high-entropy amino acid sites is dispersed, spread on the
HA1 domain. Figure S2
represents the distribution of high
entropy amino acid sites in the sequence, and Figure 3
shows part
of the high entropy amino acid sites in the 3D structure. High entropy amino acid sites can be either inside or outside the epitope
region, and they appear in every epitope. Some of these high entropy amino acid sites even have low RSA, indicating
that they are not fully exposed on the surface.

If we set a critical value of high-entropy as 0.1 \cite{deem2009epitope},
we find that some of
these high-entropy amino acid sites do not cluster in any of the epitopes
we defined. 
The fraction of high-entropy
amino acid sites in each epitope of the Victoria model are 9 out of 22 in
epitope A, 15 out of 25 in epitope B, 1 out of 23 in epitope C, 12
out of 40 in epitope D, and 8 out of 25 in epitope E, with
7 high-entropy amino acid sites outside the epitope region.
These amino acid sites are shown in Figure S2.

Re-clustering all the high-entropy amino acid sites did not improve the 
correlation between vaccine effectiveness and antigenic distance.
We analyzed sequence data from the original dataset of dimensional
reduction and clustered the behavior of high-entropy amino acid sites in
each season into three novel agglomerative clusters using Cluster
Analysis of MATLAB \cite{MATLAB:2012}. There was no significant correlation
among them and the original epitopes. Additionally, vaccine effectiveness
became zero at $p_{\rm epitope}$ or $p_{\rm all\ epitopes}$ values larger
than 1.

We also calculated the antigenic distances when all the low-entropy
amino acid sites in epitopes were removed and correlated these
antigenic distances determined from high-entropy amino acid sites only with vaccine
effectiveness. Epitope C was removed as it contained only one high-entropy
amino acid site. Therefore, in the Victoria lineage, 44 high-entropy
amino acid sites were taken into account. The linear least squares fit of $p_{{\rm epitope}}$
was $E=-0.4408\ p_{{\rm epitope}}+0.6675$, and $p_{{\rm all\ epitopes}}$
was $E=-0.5392\ p_{{\rm all\ epitopes}}+0.6735$. The estimated substitution
number of $p_{{\rm epitope}}$ was 13.03, similar to that in
Figure 1,
further substantiating the $p_{\rm epitope}$ is suitable
as the measurement criteria. However, for $p_{\rm all\ epitopes}$, vaccine effectiveness would not
decline to zero even if all these 44 high-entropy amino acid sites are substituted.
A linear least squares fit of $p_{{\rm two\ epitopes}}$ was $E=-0.6662\ p_{{\rm two\ epitopes}}+0.6704$.
This result indicates that when all of the amino acid sites in the two
dominant epitope region are mutated, vaccine effectiveness will decline
to zero. Nonetheless, these high-entropy amino acid sites contain only part
of the surface amino acid sites. The region potentially recognized by the
immune system should cover the surface of
HA1. Thus, it is likely not enough to consider only high-entropy amino acid sites only
in epitope determination. We look forward to both more epidemiological
studies for influenza B and more sequence data to extend our dataset,
in which case an updated calculation of the high-entropy amino acid sites may more fully cover the HA1
surface.

\subsection{Role of High-Entropy Amino Acid Sites with low RSA}

From Figure S2,
our definition of the five epitopes includes
most of the high entropy amino acid sites. Though we have already considered
the high-entropy amino acid sites in epitope definition, we found that there
are still some amino acid sites with average seasonal entropy greater than
0.1, such as amino acid site 262 and 267. The reason why they were not included
is that they have a low 
RSA, and they are not likely to be recognized by antibodies because they
are not on the surface of the protein.

We further tested if these amino acid sites can affect the correlation between
antigenic distance and vaccine effectiveness. Though they are not
fully exposed, substitutions in these amino acid sites could affect the epitope structure,
and substitutions of these amino acid sites could change the shape of antibody
recognition amino acid site, and affect vaccine effectiveness. We added these
7 high-entropy but low RSA amino acid sites to our 5 epitopes using the same
method as described in Section \ref{method:epitope}. In the Victoria
numbering, amino acid site 262, 266, and 267 were added to epitope B, amino acid site
29 was added to epitope C, amino acid sites 175 and 252 were added to epitope
D, and amino acid site 76 was added to epitope E. These modifications did not
quantitatively change the vaccine effectiveness and $p_{{\rm epitope}}$
relation. 

\clearpage

\begin{table}[h]
\begin{centering}
\caption{Number of sequences collected per year for the entropy analysis\label{table8} }
{\footnotesize{}}%
\begin{tabular}{llll}
  {\footnotesize{}Season} & {\footnotesize{}Victoria } & {\footnotesize{}Yamagata }
 & {\footnotesize{} total}\\
\hline 
{\footnotesize{} 1994-1995} & {\footnotesize{} 0} & {\footnotesize{} 19} & {\footnotesize{} 19} \\
{\footnotesize{} 1995-1996} & {\footnotesize{} 0} & {\footnotesize{} 6} & {\footnotesize{} 6} \\
{\footnotesize{} 1996-1997} & {\footnotesize{} 0} & {\footnotesize{} 23} & {\footnotesize{} 23} \\
{\footnotesize{} 1997-1998} & {\footnotesize{} 2} & {\footnotesize{} 0} & {\footnotesize{} 2} \\
{\footnotesize{} 1998-1999} & {\footnotesize{} 2} & {\footnotesize{} 23} & {\footnotesize{} 25} \\
{\footnotesize{} 1999-2000} & {\footnotesize{} 0} & {\footnotesize{} 29} & {\footnotesize{} 29} \\
{\footnotesize{} 2000-2001} & {\footnotesize{} 5} & {\footnotesize{} 87} & {\footnotesize{} 92} \\
{\footnotesize{} 2001-2002} & {\footnotesize{} 24} & {\footnotesize{} 19} & {\footnotesize{} 43} \\
{\footnotesize{} 2002-2003} & {\footnotesize{} 57} & {\footnotesize{} 19} & {\footnotesize{} 76} \\
{\footnotesize{} 2003-2004} & {\footnotesize{} 4} & {\footnotesize{} 33} & {\footnotesize{} 37} \\
{\footnotesize{} 2004-2005} & {\footnotesize{} 81} & {\footnotesize{} 208} & {\footnotesize{} 289} \\
{\footnotesize{} 2005-2006} & {\footnotesize{} 121} & {\footnotesize{} 41} & {\footnotesize{} 162} \\
{\footnotesize{} 2006-2007} & {\footnotesize{} 178} & {\footnotesize{} 59} & {\footnotesize{} 237} \\
{\footnotesize{} 2007-2008} & {\footnotesize{} 140} & {\footnotesize{} 365} & {\footnotesize{} 505} \\
{\footnotesize{} 2008-2009} & {\footnotesize{} 190} & {\footnotesize{} 79} & {\footnotesize{} 269} \\
{\footnotesize{} 2009-2010} & {\footnotesize{} 169} & {\footnotesize{} 29} & {\footnotesize{} 198} \\
{\footnotesize{} 2010-2011} & {\footnotesize{} 606} & {\footnotesize{} 84} & {\footnotesize{} 690} \\
{\footnotesize{} 2011-2012} & {\footnotesize{} 130} & {\footnotesize{} 133} & {\footnotesize{} 263} \\
{\footnotesize{} 2012-2013} & {\footnotesize{} 97} & {\footnotesize{} 425} & {\footnotesize{} 522} \\
{\footnotesize{} 2013-2014} & {\footnotesize{} 109} & {\footnotesize{} 185} & {\footnotesize{} 29}4 \\
\end{tabular}
\par\end{centering}{\footnotesize \par}

The number of sequences downloaded from GenBank in each September-March season
for the entropy analysis.
The sequences are classified as either Victoria or Yamagata strains.
\end{table}

\clearpage

\begin{table}[h]
\begin{centering}
\caption{Number of sequences collected per year for the dimensional reduction \label{table9} }
{\footnotesize{}}%
\begin{tabular}{llll}
  {\footnotesize{}Season} & {\footnotesize{} total} & {\footnotesize{}Season} & {\footnotesize{} total}\\
\hline 
{\footnotesize{}  1960  }  &   {\footnotesize{} 1 }& {\footnotesize{}  1988  }  &   {\footnotesize{} 7 } \\
{\footnotesize{}  1961  }  &   {\footnotesize{} 2 }& {\footnotesize{}  1989  }  &  {\footnotesize{} 14 } \\
{\footnotesize{}  1962  }  &   {\footnotesize{} 2 }& {\footnotesize{}  1990  }  &  {\footnotesize{} 31 } \\
{\footnotesize{}  1963  }  &   {\footnotesize{} 2 }& {\footnotesize{}  1991  }  &  {\footnotesize{} 14 } \\
{\footnotesize{}  1964  }  &   {\footnotesize{} 3 }& {\footnotesize{}  1992  }  &   {\footnotesize{} 8 } \\
{\footnotesize{}  1965  }  &   {\footnotesize{} 2 }& {\footnotesize{}  1993  }  &  {\footnotesize{} 20 } \\
{\footnotesize{}  1966  }  &   {\footnotesize{} 2 }& {\footnotesize{}  1994  }  &  {\footnotesize{} 23 } \\
{\footnotesize{}  1967  }  &   {\footnotesize{} 2 }& {\footnotesize{}  1995  }  &  {\footnotesize{} 28 } \\
{\footnotesize{}  1968  }  &   {\footnotesize{} 2 }& {\footnotesize{}  1996  }  &  {\footnotesize{} 29 } \\
{\footnotesize{}  1969  }  &   {\footnotesize{} 2 }& {\footnotesize{}  1997  }  &  {\footnotesize{} 70 } \\
{\footnotesize{}  1970  }  &   {\footnotesize{} 2 }& {\footnotesize{}  1998  }  &  {\footnotesize{} 26 } \\
{\footnotesize{}  1971  }  &   {\footnotesize{} 2 }& {\footnotesize{}  1999  }  &  {\footnotesize{} 64 } \\
{\footnotesize{}  1972  }  &   {\footnotesize{} 2 }& {\footnotesize{}  2000  }  &  {\footnotesize{} 35 } \\
{\footnotesize{}  1973  }  &   {\footnotesize{} 3 }& {\footnotesize{}  2001  }  & {\footnotesize{} 134 } \\
{\footnotesize{}  1974  }  &   {\footnotesize{} 4 }& {\footnotesize{}  2002  }  & {\footnotesize{} 164 } \\
{\footnotesize{}  1975  }  &   {\footnotesize{} 1 }& {\footnotesize{}  2003  }  &  {\footnotesize{} 90 } \\
{\footnotesize{}  1976  }  &   {\footnotesize{} 4 }& {\footnotesize{}  2004  }  & {\footnotesize{} 184 } \\
{\footnotesize{}  1977  }  &   {\footnotesize{} 2 }& {\footnotesize{}  2005  }  & {\footnotesize{} 345 } \\
{\footnotesize{}  1978  }  &   {\footnotesize{} 2 }& {\footnotesize{}  2006  }  & {\footnotesize{} 290 } \\
{\footnotesize{}  1979  }  &   {\footnotesize{} 2 }& {\footnotesize{}  2007  }  & {\footnotesize{} 377 } \\
{\footnotesize{}  1980  }  &   {\footnotesize{} 2 }& {\footnotesize{}  2008  }  & {\footnotesize{} 544 } \\
{\footnotesize{}  1981  }  &   {\footnotesize{} 3 }& {\footnotesize{}  2009  }  & {\footnotesize{} 336 } \\
{\footnotesize{}  1982  }  &   {\footnotesize{} 6 }& {\footnotesize{}  2010  }  & {\footnotesize{} 523 } \\
{\footnotesize{}  1983  }  &   {\footnotesize{} 4 }& {\footnotesize{}  2011  }  & {\footnotesize{} 570 } \\
{\footnotesize{}  1984  }  &   {\footnotesize{} 3 }& {\footnotesize{}  2012  }  & {\footnotesize{} 572 } \\
{\footnotesize{}  1985  }  &   {\footnotesize{} 8 }& {\footnotesize{}  2013  }  & {\footnotesize{} 419 } \\
{\footnotesize{}  1986  }  &   {\footnotesize{} 7 }& {\footnotesize{}  2014  }  & {\footnotesize{} 398 } \\
{\footnotesize{}  1987  } & {\footnotesize{}9  } &  & \\
\end{tabular}
\par\end{centering}{\footnotesize \par}

The number of sequences downloaded from GenBank in each year
for the dimensional reduction.
\end{table}

\clearpage

\begin{table}[h]
\begin{centering}
\caption{Predicted amino acid substitutions for which VE decreases to zero
\label{table3} }
{\footnotesize{}}%
\begin{tabular}{@{}llll@{}}
 & {\footnotesize{}Antigenic distances} & {\footnotesize{}Average number of total amino acids } & {\footnotesize{}Average number of substitution }\tabularnewline
 & {\footnotesize{} at zero effectiveness} &  & {\footnotesize{} at zero effectiveness}\tabularnewline
\hline 
{\footnotesize{}$p_{{\rm epitope}}$ } & {\footnotesize{}0.79 } & {\footnotesize{}24} & {\footnotesize{}19}\tabularnewline
{\footnotesize{}$p_{{\rm two\ epitopes}}$ } & {\footnotesize{}0.74 } & {\footnotesize{}51} & {\footnotesize{}38 }\tabularnewline
{\footnotesize{}$p_{{\rm all\ epitopes}}$ } & {\footnotesize{}0.43 } & {\footnotesize{}135} & {\footnotesize{}57 }\tabularnewline
{\footnotesize{}$p_{{\rm sequence}}$ } & {\footnotesize{}0.20 } & {\footnotesize{}347 } & {\footnotesize{}69 }\tabularnewline
\end{tabular}
\par\end{centering}{\footnotesize \par}

The value of antigenic distance for which the predicted effectiveness
decays to zero is shown for different measures of antigenic distance.
The number of substitutions at which the predicted effectiveness decays
to zero from each measure is also shown.
\end{table}

\clearpage

\begin{table}[h]
\begin{centering}
\caption{The dominant epitope varies by season\label{table4} }
{\footnotesize{}}%
\begin{tabular}{lllll}
\hline 
{\footnotesize{}Seasons} & {\footnotesize{}First three dominant epitopes} & $p_{{\rm dominant\ epitope}}$ & $p_{{\rm second\ dominant\ epitope}}$ & {\footnotesize{}$p_{{\rm third\ dominant\ epitope}}$}\tabularnewline
\hline 
{\footnotesize{}2012--2013 (a) } & {\footnotesize{}-} & {\footnotesize{}-} & {\footnotesize{}-} & {\footnotesize{}-}\tabularnewline
{\footnotesize{}2012--2013 (b) } & {\footnotesize{}B E D} & {\footnotesize{}0.0833333} & {\footnotesize{}0.08} & {\footnotesize{}0.05}\tabularnewline
{\footnotesize{}2012--2013 (c) } & {\footnotesize{}A B E} & {\footnotesize{}0.363636} & {\footnotesize{}0.333333} & {\footnotesize{}0.24}\tabularnewline
{\footnotesize{}2011--2012 (a) } & {\footnotesize{}B A E} & {\footnotesize{}0.4} & {\footnotesize{}0.363636} & {\footnotesize{}0.24}\tabularnewline
{\footnotesize{}2011--2012 (b) } & {\footnotesize{}A B E} & {\footnotesize{}0.363636} & {\footnotesize{}0.36} & {\footnotesize{}0.24}\tabularnewline
{\footnotesize{}2011--2012 (c) } & {\footnotesize{}-} & {\footnotesize{}-} & {\footnotesize{}-} & {\footnotesize{}-}\tabularnewline
{\footnotesize{}2011 } & {\footnotesize{}-} & {\footnotesize{}-} & {\footnotesize{}-} & {\footnotesize{}-}\tabularnewline
{\footnotesize{}2010 } & {\footnotesize{}-} & {\footnotesize{}-} & {\footnotesize{}-} & {\footnotesize{}-}\tabularnewline
{\footnotesize{}2008--2009 } & {\footnotesize{}A B E} & {\footnotesize{}0.363636 } & {\footnotesize{}0.333333} & {\footnotesize{}0.28}\tabularnewline
{\footnotesize{}2008 } & {\footnotesize{}-} & {\footnotesize{}-} & {\footnotesize{}-} & {\footnotesize{}-}\tabularnewline
{\footnotesize{}2007--2008 } & {\footnotesize{}B A E} & {\footnotesize{}0.4} & {\footnotesize{}0.363636} & {\footnotesize{}0.28}\tabularnewline
{\footnotesize{}2006--2007 } & {\footnotesize{}A B E} & {\footnotesize{}0.363636} & {\footnotesize{}0.36} & {\footnotesize{}0.24}\tabularnewline
{\footnotesize{}2005--2006 (a) } & {\footnotesize{}A B E} & {\footnotesize{}0.363636} & {\footnotesize{}0.333333} & {\footnotesize{}0.24}\tabularnewline
{\footnotesize{}2005--2006 (b) } & {\footnotesize{}A B E} & {\footnotesize{}0.454545} & {\footnotesize{}0.375} & {\footnotesize{}0.217 }\tabularnewline
{\footnotesize{}1987--1988 } & {\footnotesize{}B D A} & {\footnotesize{}0.12 } & {\footnotesize{}0.075} & {\footnotesize{}0.0454545}\tabularnewline
{\footnotesize{}1985--1986 } & {\footnotesize{}B D A} & {\footnotesize{}0.28} & {\footnotesize{}0.125} & {\footnotesize{}0.0909091}\tabularnewline
{\footnotesize{}1983--1984 } & {\footnotesize{}B D} & {\footnotesize{}0.12} & {\footnotesize{}0.025} & {\footnotesize{}0}\tabularnewline
{\footnotesize{}1979--1980 } & {\footnotesize{}B A E} & {\footnotesize{}0.12} & {\footnotesize{}0.0909091} & {\footnotesize{}0.04}\tabularnewline
\end{tabular}
\par\end{centering}{\footnotesize \par}

The first three dominant epitopes in each year in descending
order and the antigenic distances associated with each epitope.
\end{table}

\clearpage

\begin{figure}
\caption{Phylogenetic tree of representative influenza B strains\label{fig3}}

\begin{centering}
\includegraphics{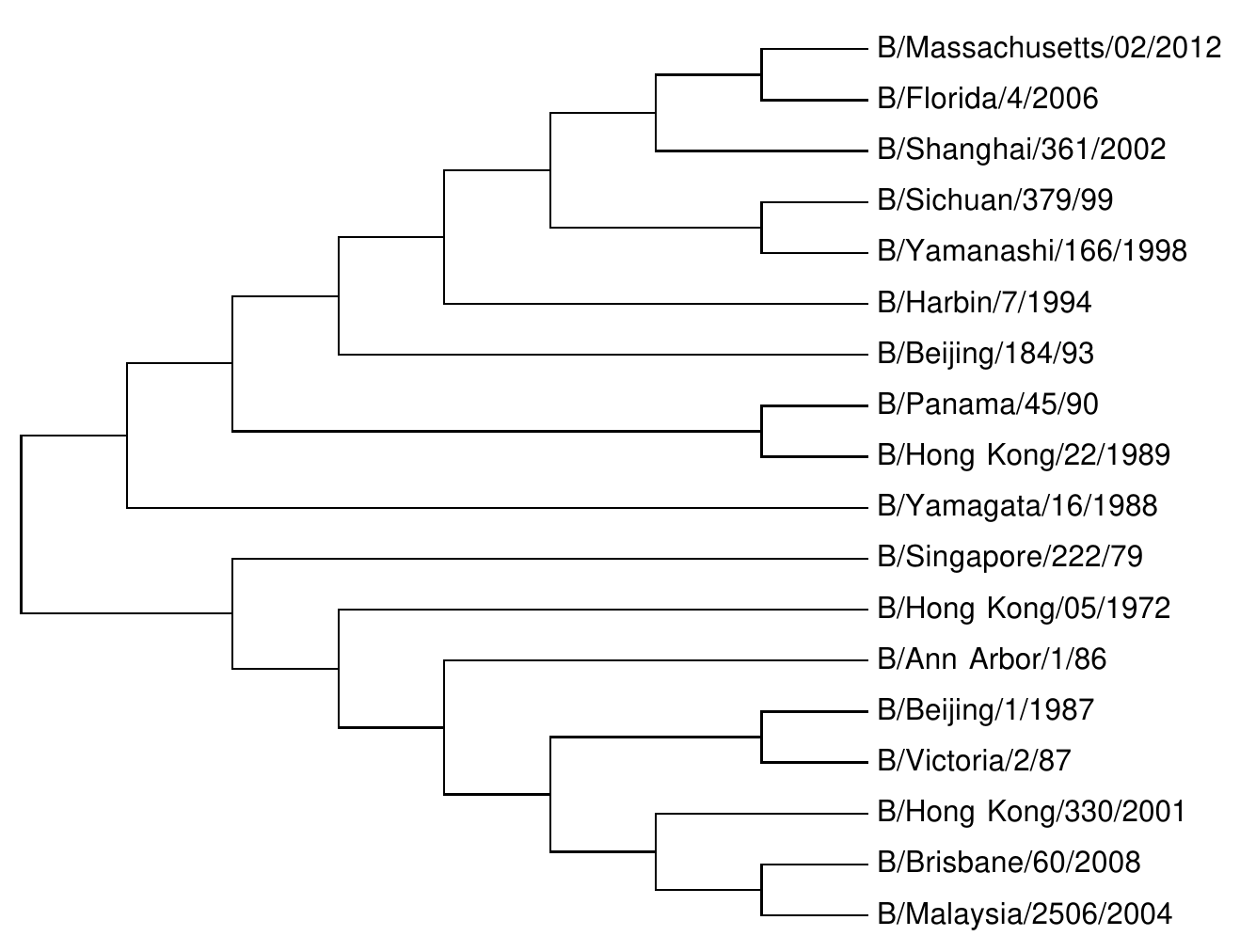}
\par\end{centering}

Maximum likelihood phylogenetic tree of the 18 representative influenza
B strains highlighted in Figure 2.
\end{figure}

\begin{figure}
\caption{Seasonal average entropy distribution in HA1 sequence\label{fig6}}

\begin{centering}
\includegraphics[scale=0.30]{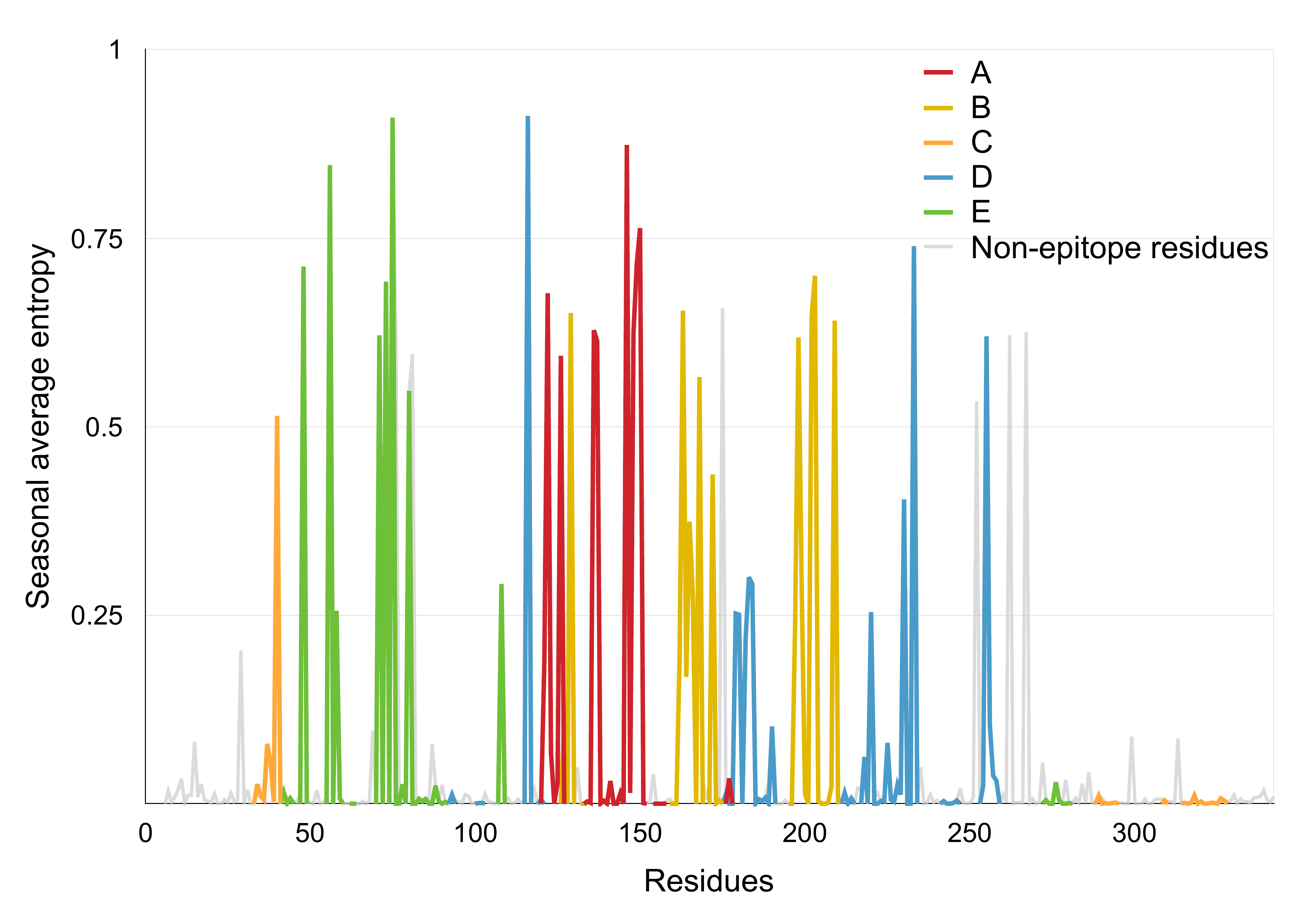}
\par\end{centering}

Seasonal average entropy distribution in HA1 sequence of influenza
B. The Victoria lineage (model 4FQM) numbering is used. Data were
collected in every Northern Hemisphere influenza season (September
to March of the next year) from 1994 to 2014, 20 seasons in total.
\end{figure}

\clearpage

\bibliography{IB_reference}